# Habitability in the Solar System and New Planetary Missions

Pauli E. Laine[1]

*1. Department of Computer Science and Information Systems, University of Jyväskylä, 40014, Finland*

**Abstract:** Definition of habitability depends on the organisms under consideration. One way to determine habitability of some environment is to compare it's certain parameters to environments where extremophilic micro-organisms thrive on Earth. We can also define more common habitability criteria from the life as we know it. These criteria include basic elements, liquid water and an energy source. We know that some locations in our Solar System provide at least some of these limits and criteria. This article describes the aims and technical specifications of some planetary missions, such as NASA's MSL in 2012, ESA's ExoMars missions in 2016 and 2018, and JUICE in 2033. These missions will explore habitability of Mars, Europa, Ganymede and Callisto. Here we compare defined habitability criteria to instrumentation documentation to determine whether these missions could validate the habitability of Mars and those Jovian moons. These missions have about 13 habitability assessment related instruments for Mars, 3 for Europa, 5 for Ganymede and Callisto. I conclude that these missions will yield important information for habitability assessment of these targets, especially Mars.

## 1. Introduction

We do not have a clear definition of life. We are not sure if life can be built differently than Earth's life. If we don't want to speculate on alien life forms, then the only logical way is to search limits where life on Earth can not only survive but also reproduce and evolve. On Earth we have found life in temperature range of 253-395 K, in pH values 0-13, in radiation up to 16 000 Gy, and in pressure conditions from near vacuum up to 142 MPa, cf. e.g. [1], [2].

More general way to define habitability is to list the most essential factors that are required for life on Earth. All basic organic molecules are built on few elements: carbon (C), hydrogen (H), nitrogen (N), oxygen (O), phosphorus (P), and sulfur (S). Liquid water is crucial for all life on Earth, as it is a unique solvent and medium for chemical reactions. Moreover all biochemical processes need some free energy to power metabolism, so some steady chemical or photoelectric energy source is needed.

All these ingredients are necessary but not sufficient for life. Life needs time and stable conditions to begin, survive and evolve. On the other hand it might be difficult or even impossible to know all possible conditions in the history of a possible currently habitable location. This means that if we want to find extant life on some place in the Solar System, we must concentrate on locations that meet today our definition of habitability. They are the first spots to start searching extraterrestrial life.

In addition to Earth, there are some suggested habitable objects in our Solar System today. The most famous of them is of course Mars. The surface of Mars is now hostile to all known life forms, but there could be subsurface locations or pockets that could have served as retreat for previous surface dwellers. Other promising objects are some moons of gas giant planets: Europa, Ganymede and Callisto of Jupiter, Titan and Enceladus of Saturn. There are indications of liquid

---

**Corresponding author:** Pauli E. Laine. E-mail: pauli.e.laine@jyu.fi.

water ocean on Europa, Ganymede, Callisto and Enceladus. Liquid water is of course flag waver to astrobiologists because it is essential to all known life on Earth. Titan is of astrobiology interest because of its presumed chemical resemblance to early prebiotic Earth [3].

## 2. Missions and Targets

There are some ongoing and planned planetary missions that have astrobiological and habitability objectives. In this paper, I will concentrate on one working, one decided and three planned missions.

NASA's Mars Science Laboratory (MSL) Curiosity landed on Mars on 6th of August 2012 and is now sending data. Shortly after MSL's successful landing, NASA announced that it had selected InSight (Interior Exploration using Seismic Investigations, Geodesy and Heat Transport) lander mission to investigate the interior of Mars in 2016.

European Space Agency's (ESA) ExoMars is a planned Mars exploration program with an orbiter/lander in 2016 and a rover in 2018. Currently an agreement between ESA and Roscosmos (the Russian space agency) has been approved, and at least the 2016 orbiter/lander mission seems likely. In April 2012 ESA announced the planning of the JUICE (JUpiter ICy moons Explorer) mission of Cosmic Vision programme.

There are currently no planned new missions to the Saturnian system (Titan and Enceladus), but NASA's Cassini probe has been exploring it since 2004. The number of instruments, their scientific goals and precision is always a compromise between other instruments, the mission profile and available payload mass. Usually missions have overarching scientific themes or goals. So MSL investigates habitability, climate and geology, InSight concentrates on geology, ExoMars has astrobiological objectives and JUICE investigates the emergence of habitable worlds around gas giants.

## 3. Instruments

How these missions and what kind of instruments can help define the assessment of the habitability of their targets? Nearly all of these missions include *spectrometers*. MSL has APXS (Alpha Particle X-ray Spectrometer) and two additional spectrometers in SAM [4]. The ExoMars Rover has ISEM (Infrared Spectrometer for ExoMars), a mast IR spectrometer; Ma_MISS (Mars Multispectral Imager for Subsurface Studies), an IR spectrometer in the rover's subsurface drill; MicrOmega (a visible plus IR imaging spectrometer); RLS (Raman Laser Spectrometer); and MOMA, a combined laser-desorption, thermal volatilisation gas-chromatograph mass spectrometer for detecting organic molecules [5]. JUICE has MAJIS (Moons and Jupiter Imaging Spectrometer), an IR spectrometer with Mars Express heritage [6].

MSL has also an X-ray *powder diffraction* instrument called CheMin (Chemistry and Mineralogy) that can do structural characterization of minerals in soil samples. Both spectrometer and X-ray powder diffraction instrument can identify possible chemical energy sources, such as hydrogen, hydrogen sulphide, and methane as electron donors and sulphate, nitrate, iron (III), and manganese (IV) as electron acceptors [7]. However, only APXS or an X-ray fluorescence instrument (like CheMin) are capable to determine atomic composition (i.e. to direct detect carbon, hydrogen, nitrogen etc.). MSL's SAM (Sample Analysis at Mars) is a suite of 3 instruments: Quadrupole Mass Spectrometer (QMS), Gas Chromatograph (GC) and Tunable Laser Spectrometer (TLS) [8]. QMS determines molecular and isotopic composition for atmospheric and evolved gas samples (crosstalk between adjacent peaks $>10^6$). *Gas chromatograph* can resolve complex mixtures of organic compounds into separate components. TLS determines abundance and isotopic composition of methane, water and carbon dioxide. For the habitability assessment, SAM will contribute for element and energy source detection.

ExoMars Rover's MOMA (Mars Organics Molecule

Analyser) is instrument for targeting biomarkers. Its spectrometer can be used together with gas chromatograph in GS-MS mode and with laser pump in LD-MS (Laser-Desorption-Mass Spectrometry) mode. Although MOMA is strictly speaking not an instrument for assessing the habitability of Mars, finding extant life will of course also mean that Mars is habitable. Alternatively, the identification of potential fossil biosignatures would indicate early Mars may have been habitable. Also LMC (Life Marker Chip) is ExoMars Rover's immunoassay instrument for biosignature detection.

Water detection can be done with a *neutron scattering* instrument, like MSL's DAN (Dynamic Albedo of Neutrons) and ExoMars Trace Gas Orbiter's FREND (Fine Resolution Epithermal Neutron Detector) [9] and Adron in the 2018 rover, or with a *radar*, like WISDOM in ExoMars (Water Ice and Subsurface Deposit Observation on Mars) in 2018 rover. Measuring depth of these instruments is however limited: about 1 meter for DAN, FREND and Adron [10] and up to 3 meters for WISDOM [11]. Also these instruments do not detect directly liquid water but rather measures distribution of H- and OH-bearing materials or subsurface ice. Detection of liquid water requires measurement of heat flow from interior (InSight's HP3 will do this in 2016); drilling to sufficient depth and using direct methods to detect possible liquid water. Question remains open whether planned ExoMars Rover's 2 meter will be enough. Of course, camera systems and spectrometers can also provide evidence of surface or near surface water and ice.

Ocean detection for Jovian icy moons is based on *magnetic field* (J-MAG) and *radio wave* (3GM, PRIDE, and RIME) measurements. Magnetic field measurement is based on moon's magnetic field's inductive response of electrical conductor, like liquid water with dissolved electrolytes [12]. Radio Doppler frequency shift in the signal between the JUICE spacecraft and ground station can reveal icy moon's tidal variation and state of their hydrostatic equilibrium, which can tell us about possible ocean [13]. JUICE's radar (RIME) studies subsurface ice structure down to 9 km, giving information about moons internal structure.

Other habitability assessment techniques include measurement of temperature, pressure (e.g. MSL/REMS), pH (e.g. MSL/SAM), and radiation (e.g. MSL/RAD). In these missions measured temperatures are surface temperatures, except InSight. pH is measured from surface soil, 2 meters below (ExoMars Rover), or remotely with spectroscopy. Radiation attenuation in soil and ice makes it negligible for habitability in these possible locations.

## 4. Mars

Tables 1 to 3 list instrumentation on each Mars mission that has some relevance to habitability assessment [4], [14], [5], [15].

| Name | Meaning | Primary objectives for habitability |
|---|---|---|
| SAM | Sample Analysis at Mars | Explores molecular and elemental chemistry relevant to life |
| APXS | Alpha Particle X-ray Spectrometer | Determines elemental chemistry |
| CheMin | Chemistry & Mineralogy | Identifies and quantifies minerals in samples |
| DAN | Dynamic Albedo of Neutrons | Detects and provides a quantitative estimation of the hydrogen in the subsurface |
| REMS | Rover Environmental Monitoring System | Explores subsurface habitability based on ground-atmosphere interaction |
| RAD | Radiation Assessment Detector | Provides input to the determination of the radiation hazard and mutagenic influences to possible life at and beneath the Martian surface |

Table 1  MSL habitability related instrumentation.

| Name | Meaning | Primary objectives for habitability |
|---|---|---|
| HP3 | Heat Flow and Physical Properties Package | Determines the temperature gradient to explore the planetary heat flow together with the physical soil properties measurements |

Table 2  InSight habitability related instrumentation.

| Name | Meaning | Primary objectives for habitability |
|---|---|---|
| NOMAD and ACS suite (2016 mission) | Nadir and Occultation for Mars Discovery | Detects trace gases and localizes sources |
| FREND (2016 mission) | Fine Resolution Epithermal Neutron Detector | Detects hydrogen on the Martian surface, targeting deposits of near-surface water ice |
| MOMA | Mars Organic Molecule Analyser | Detects organic molecules |
| RLS | Raman Laser Spectrometer | Detects mineralogical composition and identify organic pigments |
| Ma_MISS | Mars Multispectral Imager for Subsurface Studies | Images the walls of the borehole created by the drill to study Martian mineralogy and rock formation |
| WISDOM | Water Ice and Subsurface Deposit Observation On Mars | Characterises the electromagnetic properties of Martian soil in order to provide information on subsurface water content |
| Adron | Neutron detector | Detects subsurface water and hydrated minerals in combination with WISDOM |

Table 3  ExoMars habitability related instrumentation.

We gain Table 4 if we combine instrumentation information with habitability criteria. *Elements* refers to elementary building blocks of life (CHNOPS), *Liquid water* to water in form or another in search target (few meters below surface), *Energy source* means chemical elements for chemolitotrophy (e.g. $H_2$, SO, $FeS_2$, $FeCO_3$), *Temperature*, *Pressure*, *Radiation*, and *pH* refers to current limits for extremophilic microorganisms on Earth. LMC refers to Life Marker Chip onboard ExoMars Rover. Instrument(s) in parentheses refers that it can give some information about the criterion, but is not designed for that purpose.

| Criterion | Current value | MSL rover | InSight lander | ExoMars orbiter/lander | ExoMars rover |
|---|---|---|---|---|---|
| **Elements** | Mostly available, O and N equivocal | APXS, CheMin, (SAM) | - | (NOMAD) | (MOMA, RLS, Ma_MISS) |
| **Liquid water** | Possible | DAN | HP3 | FREND | WISDOM, Adron, Ma_MISS |
| **Energy source** | Oxygen equivocal | SAM, APXS, CheMin | - | - | MOMA, RLS, Ma_MISS |
| **Temperature** | 200-256 K | REMS | HP3 | Lander | - |
| **Radiation** | Negligible (underground) | RAD | - | Lander | - |
| **Pressure** | ~600 Pa (surface) | RAD/REMS | - | Lander | - |
| **Ph** | 8-9 | SAM | - | - | (LMC) |

**Table 4** Habitability matrix for Mars.

## 5. Ganymedes, Europa, and Callisto

JUICE mission's habitability assessment related instrumentation is listed in Table 5 [6], [16].

| Name | Meaning | Primary objectives for habitability |
|---|---|---|
| GALA | Ganymede Laser Altimeter | Detects the topography, shape and tidal deformation of the icy surfaces |
| RIME | Radar for Icy Moons Exploration | Obtains distributed profiling of subsurface thermal, compositional and structural horizons |
| 3GM | Gravity & Geophysics of Jupiter and Galilean Moons | Characterizes internal structure and subsurface oceans at Ganymede and Callisto and possibly at Europa |
| J-MAG | Magnetometer | Establishes and characterizes magnetic induction signatures in possible subsurface oceans |
| PRIDE | Planetary Radio Interferometer & Doppler Experiment | Characterizes internal structure and composition of Ganymede and Europa |
| MAJIS | Moons and Jupiter Imaging Spectrometer | Characterizes the composition, physical properties, geology and history of the surfaces with emphasis on the presence of organic materials, salts and weathering products |

**Table 5** JUICE habitability related instrumentation.

Ganymede is the main target for the JUICE mission. Habitability assessment related instrumentation for Ganymede is listed in Table 6 [6], [16]. It is believed, that Ganymede's ocean exists between layers of ice [12]. Inner ice layer probably cannot release enough elements for biochemistry, so Ganymede is not likely habitable [17]. JUICE mission's one scientific objective is to explore the structure of icy and ocean layers of these moons and thus clarifying their inner structure. *Temperature*, *Radiation*, *Pressure*, and *pH* here refer to possible ocean's values.

Europa is the most promising target to harbor habitable ocean and possible life, as its ocean may be in contact with the rock layer [16]. JUICE will make only two flybys, so only limited measurements are possible. Related instrumentation is listed in Table 7 [6], [16].

For Callisto there will be 12 flybys making more measurements possible. Callisto related instrumentation is listed in Table 8 [6], [16]. Callisto's possible ocean may also locate between ice layers, preventing contact with silicate floor, although inner layer seems to be composed of compressed rocks and ices, not just pure ice [18].

| Criterion | Current value | JUICE model instrument |
|---|---|---|
| Elements | No | (MAJIS, PRIDE) |
| Liquid water | Possible | GALA, RIME, J-MAG, PRIDE |
| Energy source | No | MAJIS |
| Temperature | ~250 K | RIME, PRIDE |
| Radiation | Negligible | - |
| Pressure | >100 MPa | RIME, J-MAG, PRIDE |
| pH | n/a | MAJIS |

**Table 6** Habitability matrix for Ganymede.

| Criterion | Current value | JUICE model instrument |
|---|---|---|
| Elements | Possible | (MAJIS) |
| Liquid water | Possible | RIME |
| Energy source | Possible | MAJIS |
| Temperature | 273-277 K | - |
| Radiation | Negligible | - |
| Pressure | <100 MPa | RIME |
| pH | Acidic/alkaline | MAJIS |

**Table 7** Habitability matrix for Europa.

| Criterion | Current value | JUICE model instrument |
|---|---|---|
| Elements | Possible | (MAJIS) |
| Liquid water | Possible | RIME, J-MAG, PRIDE |
| Energy source | Possible | MAJIS |
| Temperature | ~250 K | PRIDE |

| Radiation | Negligible | - |
|---|---|---|
| Pressure | >100 Mpa | RIME, J-MAG, PRIDE |
| pH | n/a | MAJIS |

**Table 8** Habitability matrix for Callisto.

## 8. Discussion

MSL has already provided first results from Mars. SAM and CheMin instruments have identified at least oxygen, sulfur dioxide and carbon dioxide from soil samples [19]. DAN has measured water equivalent hydrogen content to be about 3.9% at depth of 21 cm [20]. Liquid water remains open question, but not much is needed. Thin liquid water films would permit micro-organisms to proliferate [21]. Hygroscopic salts like perchlorate absorb moisture and could create such thin water films at Mars atmospheric conditions [22]. For underground temperature profile the heat flow need to be measured. That is what InSight will do in 2016. MSL's SAM has also found possible chemical energy source (sulfates and sulfides) for micro-organisms [19]. Initial results look promising for Mars.

JUICE instruments were selected in February 2013 [6]. Main astrobiological objective is to find proof of habitable oceans on target moons. There is evidence that such oceans are not necessary isolated but interact with the surface [23]. This means that ocean composition (elements and possible energy sources) could be detected using spectroscopic imaging, like the MAJIS instrument.

## 8. Conclusions

We have seen that all these missions will contribute to the habitability assessment of these objects. Results from different instruments will validate, weaken or even falsify (e.g. ocean ruled out) habitability hypothesis of these objects. Another question is whether these missions can actually reach potential habitable locations. For Mars, possible habitable locations are underground, and it is not yet known, what minimum depth should be examined. Neither can we yet directly reach possible ocean under kilometers thick ice on a distant moon. Fortunately oceans can be detected indirectly and their qualities can possibly be examined from the interactions with the surface.

## Acknowledgments

The author thank Axel Brandenburg (Nordita), Kirsi Lehto and Harry Lehto (University of Turku) and Jorge Vago (ESA) for discussions.
.